\documentclass[final,5p,times,authoryear]{elsarticle}

\usepackage{amssymb}
\usepackage{amsmath}
\usepackage[authoryear]{natbib}
\usepackage{siunitx}
\usepackage{textcomp}
\usepackage{color}
\usepackage{tabularx,ragged2e,booktabs,caption}
\usepackage{hyperref}

\journal{NDT \& E International }

\begin{document}

\begin{frontmatter}



\title{Augmented Ultrasonic Data for Machine Learning}

\author[label1]{Virkkunen, I.}
\author[label2]{Koskinen, T.}
\author[label2]{Jessen-Juhler, O.}
\author[label2]{ Rinta-aho, J.}

\address[label1]{Aalto University}
\address[label2]{VTT Technical Research Centre of Finland Ltd}

\begin{abstract}
Flaw detection in non-destructive testing, especially in complex signals like ultrasonic data, has thus far relied heavily on the expertise and judgement of trained human inspectors. While automated systems have been used for a long time, these have mostly been limited to using simple decision automation, such as signal amplitude threshold.

    The recent advances in various machine learning algorithms have solved many similarly difficult classification problems, that have previously been considered intractable. For non-destructive testing, encouraging results have already been reported in the open literature, but the use of machine learning is still very limited in NDT applications in the field. Key issue hindering their use, is the limited availability of representative flawed data-sets to be used for training.

In the present paper, we develop modern, very deep convolutional network to detect flaws from phased-array ultrasonic data. We make extensive use of data augmentation to enhance the initially limited raw data and to aid learning. The data augmentation utilizes virtual flaws - a technique, that has successfully been used in training human inspectors and is soon to be used in nuclear inspection qualification. The results from the machine learning classifier are compared to human performance. We show, that using sophisticated data augmentation, modern deep learning networks can be trained to achieve superhuman performance by significant margin.

\end{abstract}


\end{frontmatter}






\section[Introduction]{Introduction}

Automated systems have long been used for flaw detection in various Non-destructive evaluation (NDE) systems. The automated systems provide consistent results and do not show the variation commonly seen in human inspectors due to fatigue, stress or other factors. However, the traditional automated systems have relied on simple decision algorithms such as a signal amplitude threshold. In more demanding inspection cases, such as the typical ultrasonid inspections, the human inspectors achieve far superior inspection results than the simplistic automated systems. Consequently, in most of these inspections the data analysis are currently analyzed by human experts, even when the data acquisition is highly automated. Such analysis is time consuming to do and taxing for the personnel.

The key problem with more sophisticated automation has been, that the work of the human inspector does not lend itself to simple algorithmic description. The inspectors acquire their skill through years of training and utilize various signal characteristics in their judgement (e.g. the “signal dynamics”). Machine learning (ML) systems can be used to automate systems, where direct algorithmic description is intractable. The recent improvements in ML algorithms and computational tools (GPU acceleration, in particular) have enabled more complex and powerful models that reach near human-level performance in tasks like image classification and machine translation.

    Early attemps to use machine learning for NDT flaw detection and classification focused on using simple neural networks to classify various types of NDT data. \citet{Masnata1996} used a neural network with single hidden layer to classify various flaw types (cracks, slag inclusions, porosity) from ultrasonic A-scans. Before learning, the A-scan was reduced to 24 pre-selected features  using the Fischer discriminant analysis. \citet{Chen1993} used wavelet decomposition, to obtain features from A-scans and reported goal classification, while the training and testing was done with limited data set. \citet{Yi1998} similarly used shallow neural network to train flaw type classifier with a larger data set. Although in many cases this early work reported high classification accuracy, the results proved to be difficult to scale and to extend to new cases.

    One of the issues with developing ML-models for defect classification has been the limited availability of training data. \citet{Liu2002} used finite element simulation results to provide artificial NDT signals to augment training data.

    With the increase in computational power, the used machine learning models have become more powerful. Many authors have reported good results with shallow models like support vector machines (SVM's). While these models offer high classification capability, they also require a pre-selected set of features to be extracted from the raw NDT signal. \citet{Fei2006} used wavelet packet decomposition of ultrasonic A-scans to train SVM for defect classification in a petroleum pipeline. \cite{Sambath2010} used neural network with two hidden layers to classify ultrasonic A-scans using a hand-engineered set of 12 features. \cite{Shipway2019} used random forests to detect cracks from fluorescent penetrant inspections (FPI). \citet{Cruz2017} used feature extraction based on principal component analysis to train a shallow neural network to detect cracks from ultrasonic A-scans. He reported good classification analysis with only 5 extracted features, and computational efficiency that makes such classification feasible as on-line evaluation support for inspector during manual scanning.

\citet{Kahrobaee2018} demonstrated the use of machine learning to achieve data fusion by learning separate classification networks from different NDT data and using a combined classifier with the results from these separate classifiers. It is often the case in inspection, that more than one inspection method is used. Ability to take better advantage of the multiple data sources would thus be advantageous. Also, such approach could be used to discriminate between different flaw types, especially when the training data is too limited or separate to allow direct learning of classifiers to separate between similar flaw types.

    The machine learning classifiers have been used to wide variety of NDT signals and classification cases. \citet{Tong2018} used deep convolutional neural networks (CNNs) to detect subgrade defect from ground penetrating radar signals. For NDT methods, that provide image or image-like raw data, deep CNNs used for image classification have been applied with little modification. \citet{Dorafshan2018} used the AlexNet \citep{AlexNet} deep CNN for detecting cracks in concrete from visual inspection images.


Convolutional networks have recently shown great success with various image classification tasks \citep{marcus2018deep}. The convolutional architechture lets the networks to learn position independent classification. The recent deep architerctures have shown the ability to learn increasingly abstract representations in higher layers, which obliviates the need for hand-engineered features \citep{vgg16}. These features make the deep convolutional networks also interesting for the flaw detection in NDE signals.

Recently \citet{Meng2017}, \citet{Zhu2019} and \citet{Munir2018} used deep CNNs for defect classification in ultrasonic and EC-data. \citet{Meng2017} used deep neural networks with an SVM top layer for enhanced classification capability. The classifier was used to classify voids and delamination flaws in carbon fibre composite material. Before presented to the CNN, the raw A-scan data was decomposed using wavelet packet decomposition and the resulting coefficients re-organized into 32x16 feature matrix. Thus, the CNNs classified the A-scans separately.

    \citet{Munir2018b} used deep CNN's to classify austenitic stainless steel welds. The training data was obtained from weld training samples containing artificial flaws (i.e. solidification flaws). The data-set was augmented by shifting the A-scans in time-domain and by introducing Gaussian noise to the signal.

    \citet{Zhu2019} used deep CNN's to detect cracks in eddy current signal. Also, drop-out layer was used to estimate the confidence of the classification, which is an important opportunity in using ML in field NDT, where the reliability requirements are very high. This work is also notable in that the raw signal database was exceptionally representative with NDT indications representing plant data for various defect types \citep{epriSG}.

In summary, the current state of the art for using machine learning in NDT classification may be seen to focus on two distinct aims. Firstly, modern shallow ML models (e.g. random forests) with advanced feature-engineering are used with the aim to develop computationally lightweight models that can be implemented on-line to aid inspector in manual inspection. Secondly, deep CNNs are used to learn from raw NDT signals without the need for explicit feature engineering. The recent work on deep models takes full advantage of recent advances in models developed for other industries and shows good results across different NDT fields.

For ultrasonic testing, the existing machine learning models have mostly involved classification the single A-scan level. This is a natural approach for many applications, such as the previously studied manual inspection \citep{Cruz2017} or for C-scan style classification of large inspection analysis as done by \citet{Meng2017}. However, in many inspection cases, mechanized inspection and electronic scanning using phased array ultrasonic systems provide rich data-set where adjacent A-scans can be analysed together to provide more information. Machine learning application to such data-sets have not been widely published. In the present work, we present application of deep CNN for phased array ultrasonic data, where number of adjacent A-scans are considered together for improved flaw detection capability.

Common obstacle for using powerful ML models in NDE classification is, that the available flawed data tends to be scarce. Acquiring sufficient representative data-set would in many cases necessitate artificially manufacturing large set of flawed samples, which quickly becomes infeasible. Data augmentation is commonly considered a key tool for successful application of ML for small data sets and some authors have used data augmentation \citep{Munir2018b} for ultrasonic data. In the present work, we significantly expand on the previously published data augmentation schemes for ultrasonic inspection by using virtual flaws to generate augmented data sets. The use of virtual flaws enables generation of highly representative augmented data set for ML applications.

Finally, the key requirement for adaptation of ML machine learning models in many industries, is to show how they compare with human inspectors. Especially in high-reliability industries like the nuclear and aerospace industries, there's common requirement to employ best-available means to guarantee structural reliability. In practice, this would mean that the ML models would need to show performance exceeding that attained by the human inspectors or to show performance that meets the current requirements set for the traditional inspection systems (e.g. show required $a_{90/95}$ performance, as commonly required in the aerospace industry). However, in many cases even the human inspection performance is not quantified and known with sufficient reliability to allow direct comparison to developed ML models. In present work, we used human performance data obtained from previous research \citep{IVfeedback} and developed the machine learning models to work on comparable data thus enabling direct comparison between human inspector and modern machine learning model.

    \subsection{Virtual flaws and data augmentation}
The problem with ultrasonic training of machine learning models is the scarcity of representative ultrasonic data. Samples with real flaws are difficult to come by and in terms of nuclear power plants can be contaminated making them challenging to use. Mock-ups can be made with representative flaws, but production of such mock-ups is costly and time-consuming. The mock-ups also tend to be specific to a certain inspection case. Virtual flaws can be used to generate sufficient representative flawed ultrasonic data from limited set of mock-ups and flaws \citep{IVeFlawFirst,  SKBeFlaw,SQCeFlaw,TKinsight}. In essence, the flawed sample is scanned and the ultrasonic data recorded. From the recorded data the flaw signals are extracted by comparing the signal data point by data point to a selected flawless area. The flaw signal extracted this way is guaranteed to be representative, since it is recorded from an existing flaw. The extracted flaw signal can then be implanted into different locations of the scan data, point by point, allowing the generation of new virtual flaws. In addition, the depth and length of the flaw can be altered and various other signal modifications can be achieved. The flaw signals extracted can be moved to different samples. Flaw signals acquired with different ultrasonic parameters can be made compatible with different files.
 Using the virtual flaws augmented data generation is virtually unlimited and ample representative training data can be generated for the training of ML models. The approach has some similarity with synthesized learning cases used by \citet{bansal2018chauffeurnet}.

    \subsection{Estimation of NDT performance and probability of detection (POD) }

NDE is most valuable when used in area, where its expected reliability is very high. Consequently, measuring the performance of an NDE system and its reliability, in particular, is demanding. Demonstrating this high reliability requires high number of evaluation results on relevant targets and, thus, high number of test samples with representative flaws. Providing these flawed test samples is costly and thus different methodologies have evolved to optimize the use of the available test blocks.

Currently, the standard way to measure NDE performance is to define a probability of detection (POD) curve and, in particular the smallest crack that can be found at level of sufficient confidence, typically 90\% POD at 95\% confidence ($a_{90/95}$). Experimentally, the POD curve is determined with test block trials and a set of standardized statistical tools \citep{milhdbk, ASTMhitmiss,ASTMahat}.
\citet{}.

In this paper hit/miss method was selected due to nature of the test set-up. While signal amplitude can be used with fewer test blocks, it does not include the effects of inspector judgement on the NDE reliability. Especially in noisy inspection cases such as austenitic stainless steel welds, flaw detection relies on pattern recognition, not just signal amplitude and a clear threshold, thus the result is filtered by the inspector. This was observed also by \citet{e189ee8d126e485fb65cbd36710d911b}. For the present study and comparing human and machine inspectors, it's vital to include the judgement effect and thus, the hit/miss approach was chosen.

    \section{Materials \& Methods}
    \subsection{NDT Data} \label{NDT_Data}

\begin{figure}
	\includegraphics[width=\linewidth]{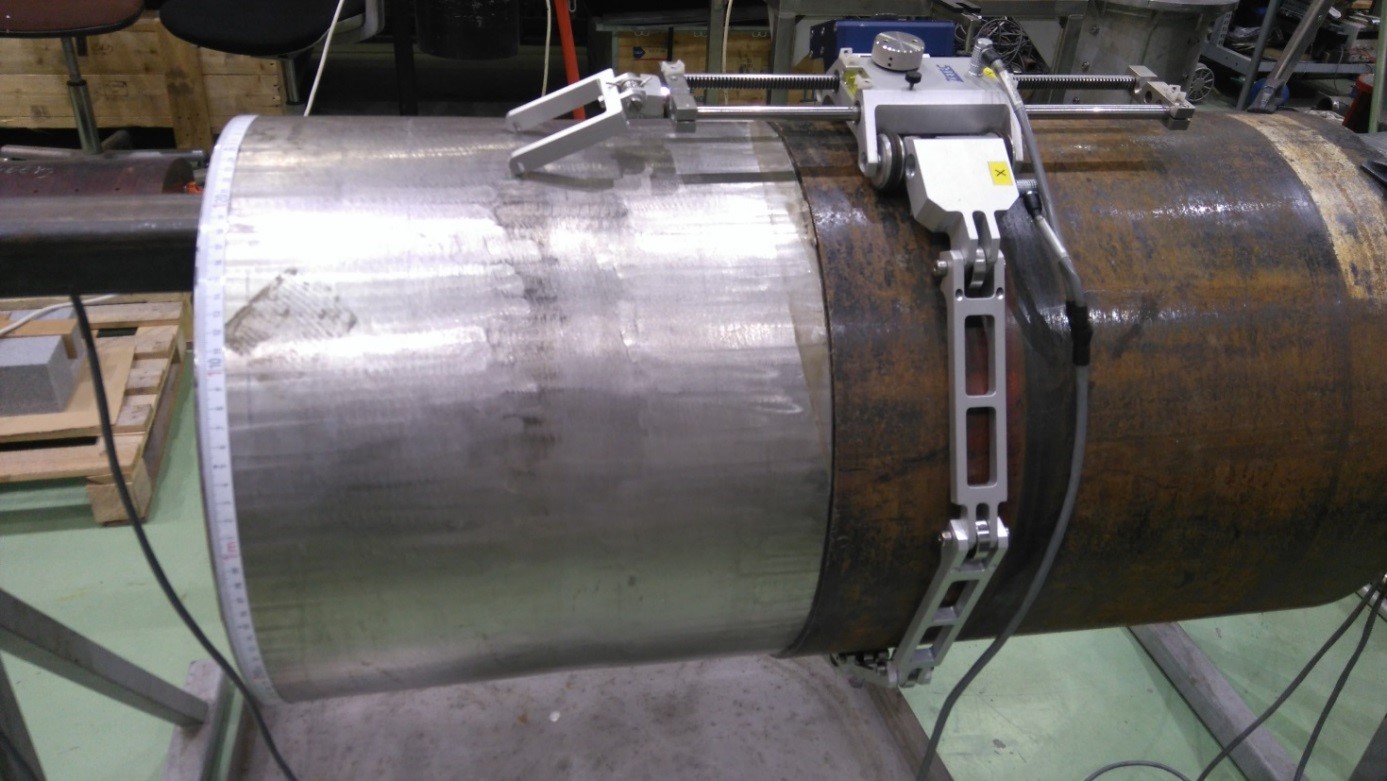}
	\caption{Scan set-up with Zetec pipe scanner, extension fixed to the right side for scanner mounting.}
	\label{fig:scan_setup}
\end{figure}

Inspected specimen for data-acquisition was a butt-weld in an austenitic 316L stainless steel pipe. Three thermal fatigue cracks with depths 1.6, 4.0 and 8.6 mm were implemented in the inner diameter of the pipe near the weld root by Trueflaw ltd. and scanned with ultrasonic equipment. An austenitic weld was chosen as a test specimen due to being common in the industry. In addition austenitic weld has increased inspection difficulty due to noise caused by the anisotropy of the weld structure.
Inspection method used for data acquisition was Transmission Receive Shear (TRS) phased array, one of the common methods used in inspecting of austenitic and dissimilar metal welds. The scan was carried out by using Zetec Dynaray 64/64PR-Lite flaw detector linked to a PC. The probes used were a Imasonic 1.5 MHz 1.5M5x3E17.5-9 matrix probes with central frequency at 1.8 MHz, element dimensions 3.35 x 2.85 mm and element arrangement as 5 x 3 elements. Used wedge was ADUX577A used to produce a shear wave efficiently. One linear scan with no skew angles was utilized. The ultrasonic wave was focused to the inner surface of the pipe and the probe was positioned in a way that the beam would be focused directly to the manufactured cracks. Coupling was applied through a feed water system and the pipe was rotated underneath the probe to assure constant and even coupling between the probe and the pipe. Probe position was carefully monitored along the scan line by Zetec pipe scanner with 0.21 mm scan resolution. The specimen and the inspection procedure is described in more detail in \cite{TKinsight}. The specimen and the scanner can be seen in Figure \ref{fig:scan_setup}.

    For data efficiency, only a signle angle was used. The chosen angle was the one, where the cracks were the most visible. In this case, this was the 45° angle. As only one scan line was acquired, the data was visualized and evaluated using B-scan images. Since the crack locations and sizes were precisely know, the crack signals could be removed from the ultrasonic data to create a blank canvas. Virtual flaw augmentation was used to broaden the representative sizes of the cracks. The virtual flaw software used was Trueflaw's eFlaw. In this case, the eFlaw  was used with an assumption that signal amplitude is the most significant feature of the crack signal from detection point of view. Similar assumption is used in the signal response POD estimation ($\hat{a}$ vs. $a$). The eFlaw was used to modify and scale down the original crack signal amplitude to represent different variety of cracks with smaller sizes than the original. This allows creation of high amount of crack images required for POD estimation and for teaching datasets for ML algorithms. Details of the eFlaw technology are explained in \citet{IVeFlawFirst,  SKBeFlaw,SQCeFlaw,TKinsight}.


The teaching data set was created in same way as for testing data set for human inspectors in previous paper \citet{IVfeedback}. Once the teaching was finished, the ML algorithm was tested with the same data as human inspectors faced. Thus, the ML algorithm and human inspectors were given the exact same information with the same controlled environment and a POD curve was estimated based on the hit/miss results.

    \subsection{Training data and used data augmentation}
The single 45\textdegree scan line data containing signals from three manufactured thermal fatigue flaws was taken as the source data for training the machine learning model. This is the same data, that was used to generate human POD results in \citep{IVfeedback}. From this data, large number of data files were generated using the same algorithm as previously. The data contained 454 A-scans each containing 5058 samples with 16 bit depth.

For machine learning purposes, the data was further processed, as follows; each A-scan was cut so that only the interesting area around the weld was included resulting in 454 x 454 point data. Then, the resolution of the ultrasonic data was down sampled to 256 x 256 points.

Altogether 20000 variations were generated to be used as training and validation data. The data was stored in minibatches of 100 UT-images per file with accompanying true state information showing the included crack state present, if any. The data set also contained information, where virtual flaw process had been used to copy unflawed section to another location. This was done to avoid and to detect the possibility that the machine learning model would learn to notice the virtual flaw introduction process, instead of the actual flaws.

    \subsection{Used ML architecture}

\begin{figure}
	\includegraphics[width=\linewidth]{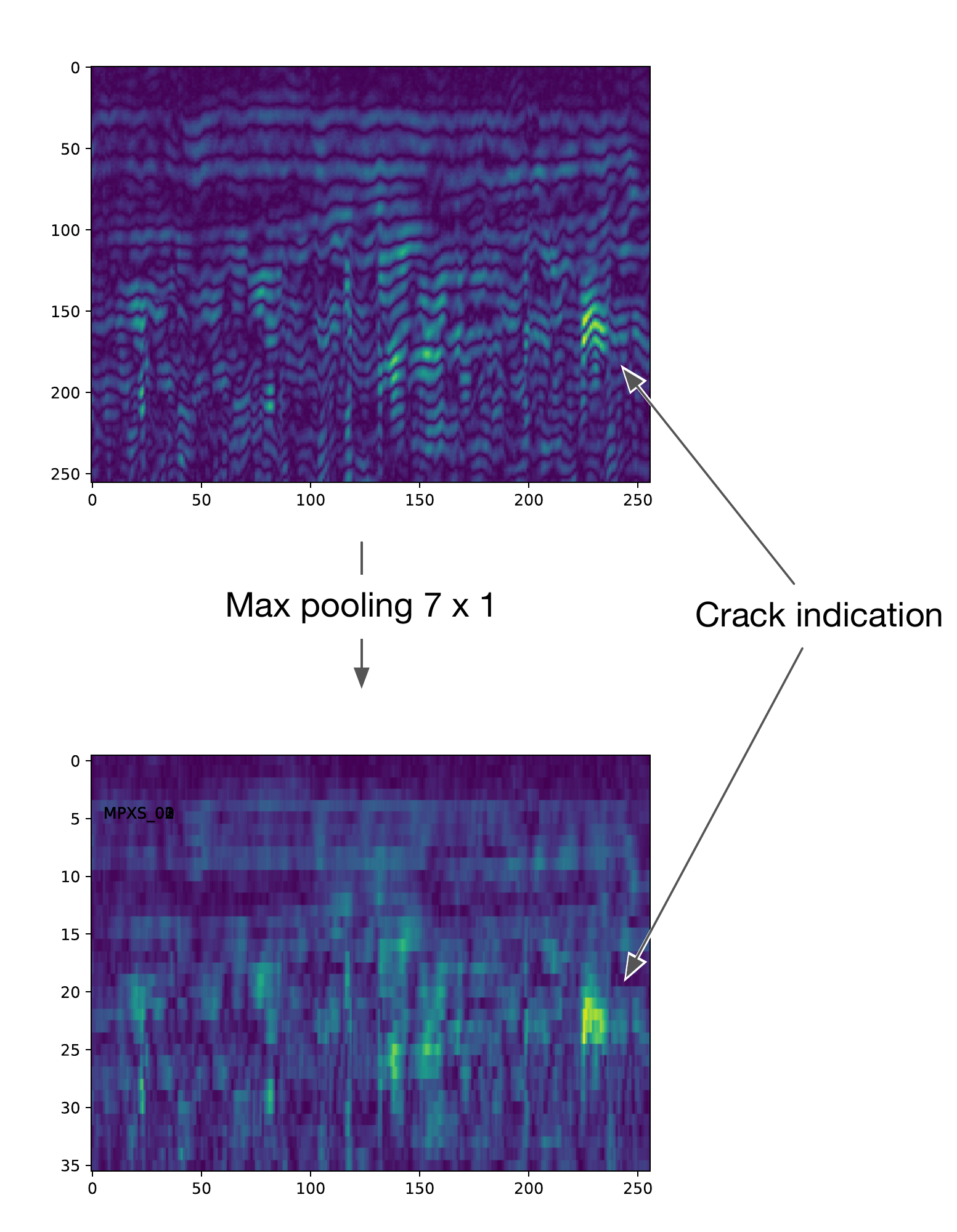}
	\caption{Max pooling was implemented as a first layer that removed the spectral informmation and reduced dimensionality of the data.}
	\label{fig:firstmaxpool}
\end{figure}

The machine learning architecture used was based on the VGG16 network \citep{vgg16}. For ultrasonic data analysis, the basic network was augmented with a first max-pooling layer, with pooling size adjusted to the wavelength of the ultrasonic signal. This max-pooling layer had the effect of removing spectral information from the image so that the rest of the network was left with an envelope amplitude curve. The effect of this layer is shown in Figure \ref{fig:firstmaxpool}. The training used binary cross entropy as the cost function and training was done using the RMSProp \citep{zeiler2012adadelta}.

Previous work \citep{Chen1993,Fei2006,Meng2017} typically extracted additional information from the spectral content of the A-scan data using, e.g., the wavelet decomposition. In this case, it was also considered to add additional data layers obtained with wavelet decomposition. However, the source data that was used for human inspectors was rectified, which made obtaining any useful information from the spectral content impossible. Since in this case, it was desirable to use data, that was directly comparable to the data seen by the human inspectors it was decided to continue working with the rectified data.

The data was read in the saved mini-batches, converted to 32 bit floating point numbers and normalized by subtracting the mean and dividing by standard deviation. A small value of 0.00001 was added to avoid division by zero.

\begin{figure}
	\includegraphics[width=\linewidth]{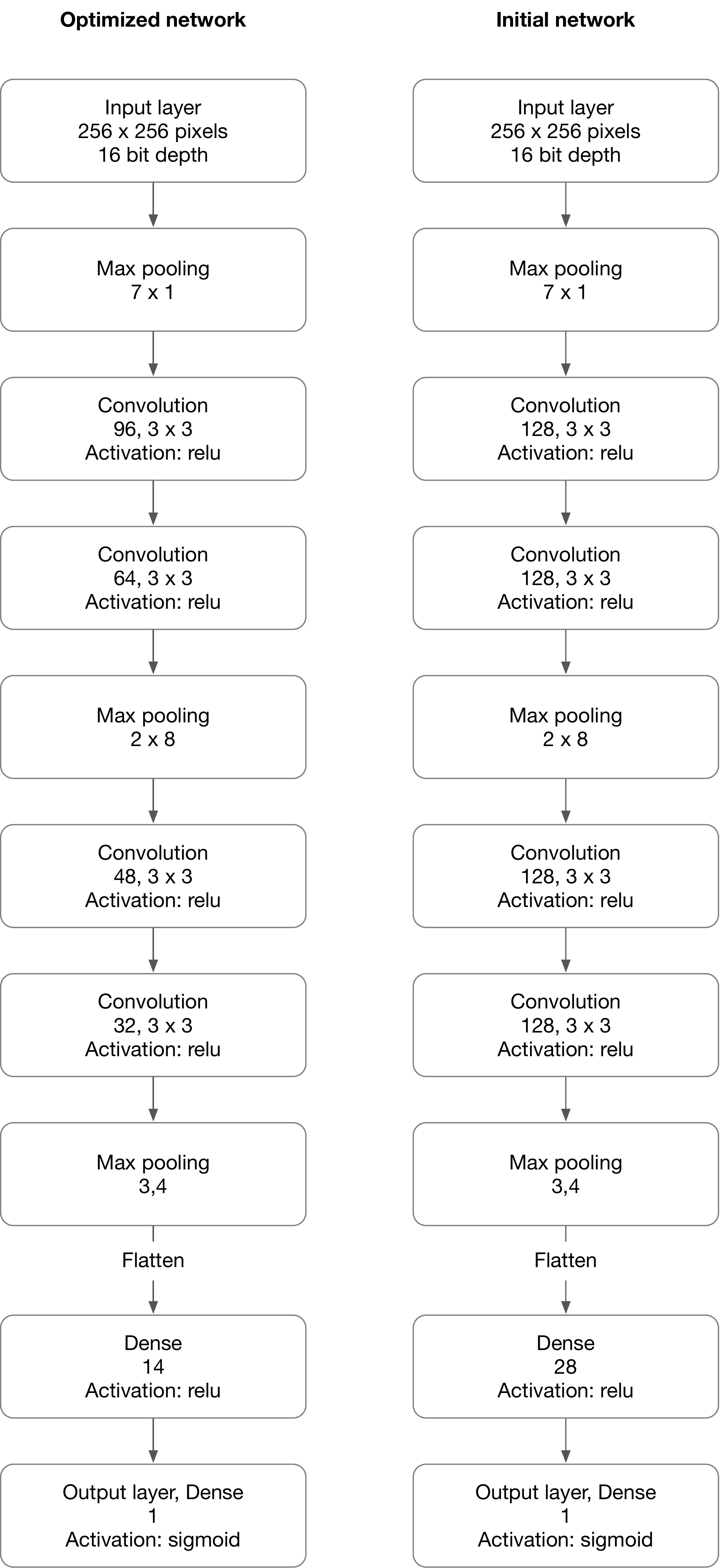}
	\caption{The trained network stucture. Max pooling was implemented using Keras MaxPooling2D layer. Convolution layers were implemented using Keras Conv2D layer. The final dense layer was implemented with Keras Dense layer.}
	\label{fig:network}
\end{figure}

The size of the various layers were originally excessive, and as soon as successful training was obtained, the layer sizes were decreased step-by-step to obtain the most efficient network capable of learning to classify the data. The full architecture (both initial trial and final) is shown in Figure \ref{fig:network}. The network experienced some sensitivity to initialization, and on repeated training, the model sometimes failed to learn successfully.

The computation was implemented with the Keras library \citep{chollet2015keras} using the TensorFlow back-end \citep{tensorflow2015-whitepaper}.

The chosen architecture does not make use of some of the recent features included in state of the art deep convolutional networks. The primary motivation for this was to keep the network as simple as possible while showing good flaw detection capability. Some of the considered, but not included, ML architectural features are discussed in the following.

Drop-out \citep{dropout} has been extensively used to prevent overfitting, and more recently to estimate prediction confidence \citep{Zhu2019}. In the present study, the model did not show susceptibility to overfitting. The likely reason for this is the high number of augmented images used for training. Consequently, drop-out was not included and instead the training was stopped after sufficient performance was achieved. Training with smaller augmented data-sets could show overfitting and, consequently, make use of drop-out. Furthermore, even in the absense of overfitting, the use of drop-out to estimate prediction accuracy is an interesting option especially in case where multiple flaw types are classified within one model.

Batch renormalization has shown to improve trainability of very deep networks \citep{renormalize}. While the present network did show sensitivity to initialization values and sometimes failed to train successfully, this did not present significant problem in this application. A simple re-try with different random starting values quickly resulted in successful training result.

Channel-wise training \citep{channelwise} has been used to ease training and to improve training results in image classification. In the present case, the interesting channel-wise information would be amplitude information (as used in the present analysis) and frequency-related information, such as the wavelet decomposed features used, e.g., by \citet{Chen1993, Fei2006}. However, in this case, it was of interest to use as-is the data that was used in previous research \citep{IVfeedback} to estimate human POD performance. As this data was rectified, most of the spectral data was lost and could not be used. Extracting spectral features using wavelet decomposition as separate channels remains interesting option for further study and may improve flaw detection.

    \subsection{Performance evaluation}
    \label{sec:performance}

In previous research \citep{IVfeedback} an online tool for assessing inspector performance was developed. The tool presents randomly generated B-scan data with implemented virtual cracks and a possibility to change the software gain. In the normal mode the inspectors select the locations of the cracks and move on to the next image. In the learning mode feedback from the previous image is provided before moving to the next image. The tool is publicly available at \url{http://www.trueflaw.com/truepod} and \url{http://www.trueflaw.com/truelearnpod}.  Not all images include cracks. The results are used to produce hit and miss POD-curve. In previous research, nine level-III ultrasonic inspection course attendees were randomly split into two groups to use the learning mode and the normal mode. Each inspector had time to practise with the tool during the course. Finally each inspector analysed 150 images and hit and miss POD-curve was generated. One inspector was excluded from the data due to excessive amount of false calls. For inspectors the best achieved $a_{90/95}$ value was at 1 mm and under 20 false calls. Most inspectors rated between 1 - 2.5 mm $a_{90/95}$ and under 30 false calls. The lower-end inspectore got $a_{90/95}$ between 3.5 and 4.0 mm and the highest false call rates were above 180. The number of false calls did not correlate with inspection performance. While the online tool does not reflect realistic inspection situation, it allows relatively rapid and cost-efficient gathering of relevant performance data. Inspection is often done in suboptimal conditions, and requires skilled inspector. In addition, the rate at which flaws appear is low making the already repetitive work even more tiring.

    The target in this study is to assess the performance of the ML model with regard to inspector performance. In addition to the previous data, that included independent inspectors, a new data set was generated. To get direct comparison between the human inspectors and the ML model, a new set of 200 B-scan images not used in the training of the ML model was generated and a hit and miss POD-curve made for the ML model. A specialized version of the previously used online tool for POD evaluation was created with this data set. Human results were then obtained from 3 experienced inspectors from VTT. The same data-set was then given to the classifier network. This set-up enabled direct comparison of human and machine performance in a blind set-up. This data-set contained 200 images and 86 images with cracks. Both the humans and the ML-network had opportunity to train with similar data and similar set-up. For these data, the range of available inspectors is more limited, but the data is even more comparable.

    \section{Results}
    \subsection{Training results}
\begin{figure}
	\includegraphics[width=\linewidth]{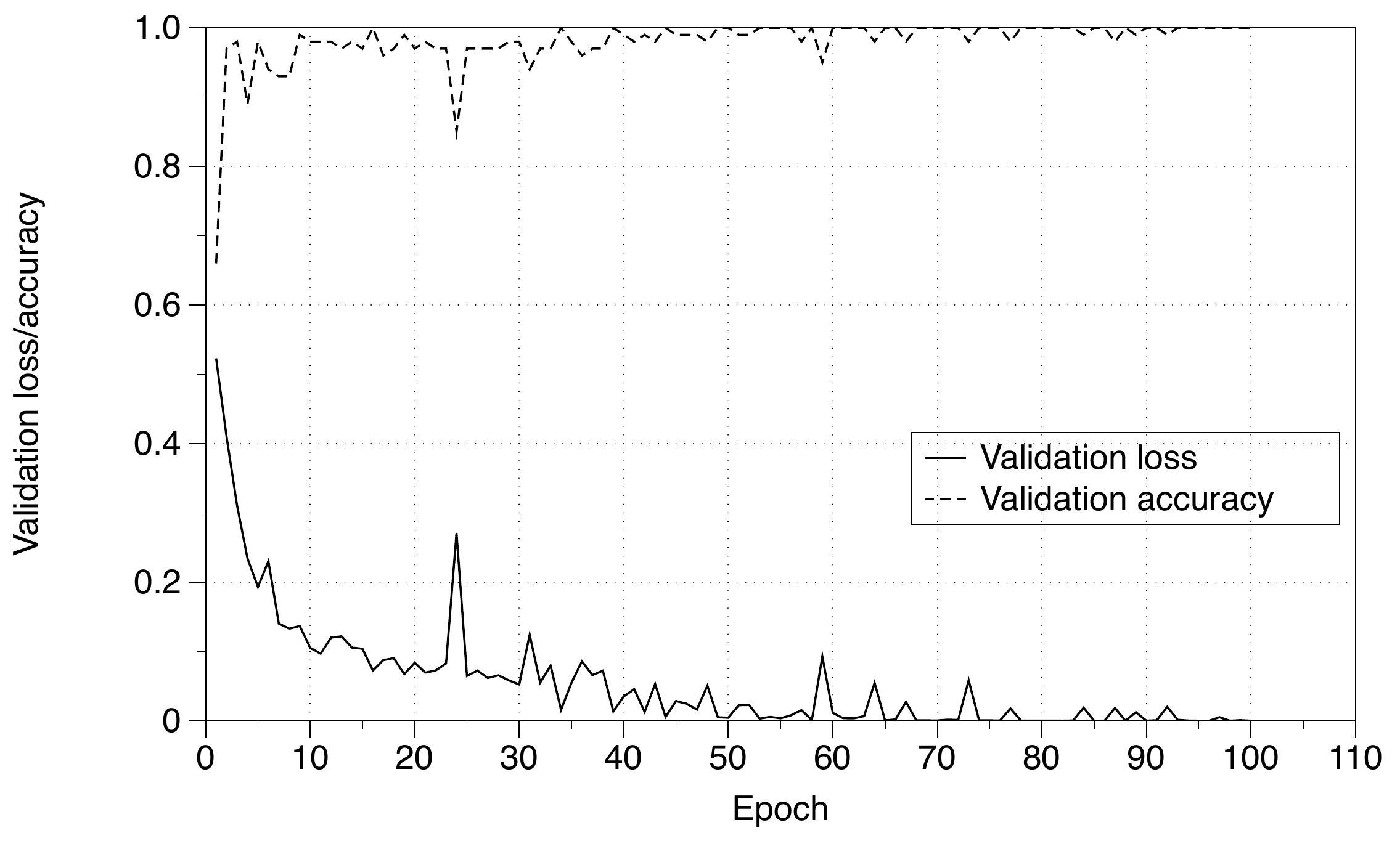}
	\caption{Validation loss and validation accuracy during training for 100 epochs.}
	\label{fig:convergence}
\end{figure}

The network was trained for 100 epochs of 10000 samples. This resulted in perfect classification: all cracks were correctly classified and no false calls were made. The evolution of the training accuracy is shown in Figure \ref{fig:convergence}. The number of training epochs was set by hand to stop slightly after perfect classification score was achieved. During development, the results were evaluated against a separate validation set. The final result was then evaluated against a previously unseen verification set. Each set contained a 100 images, with roughly 50\% cracks.

    \subsection{Comparison with human performance}

To evaluate the network performance against human performance, data set from previous work was utilized \citet{IVfeedback}. In addition a new data set was generated for this purpose (section \ref{sec:performance}).

\begin{figure}
	\includegraphics[width=\linewidth]{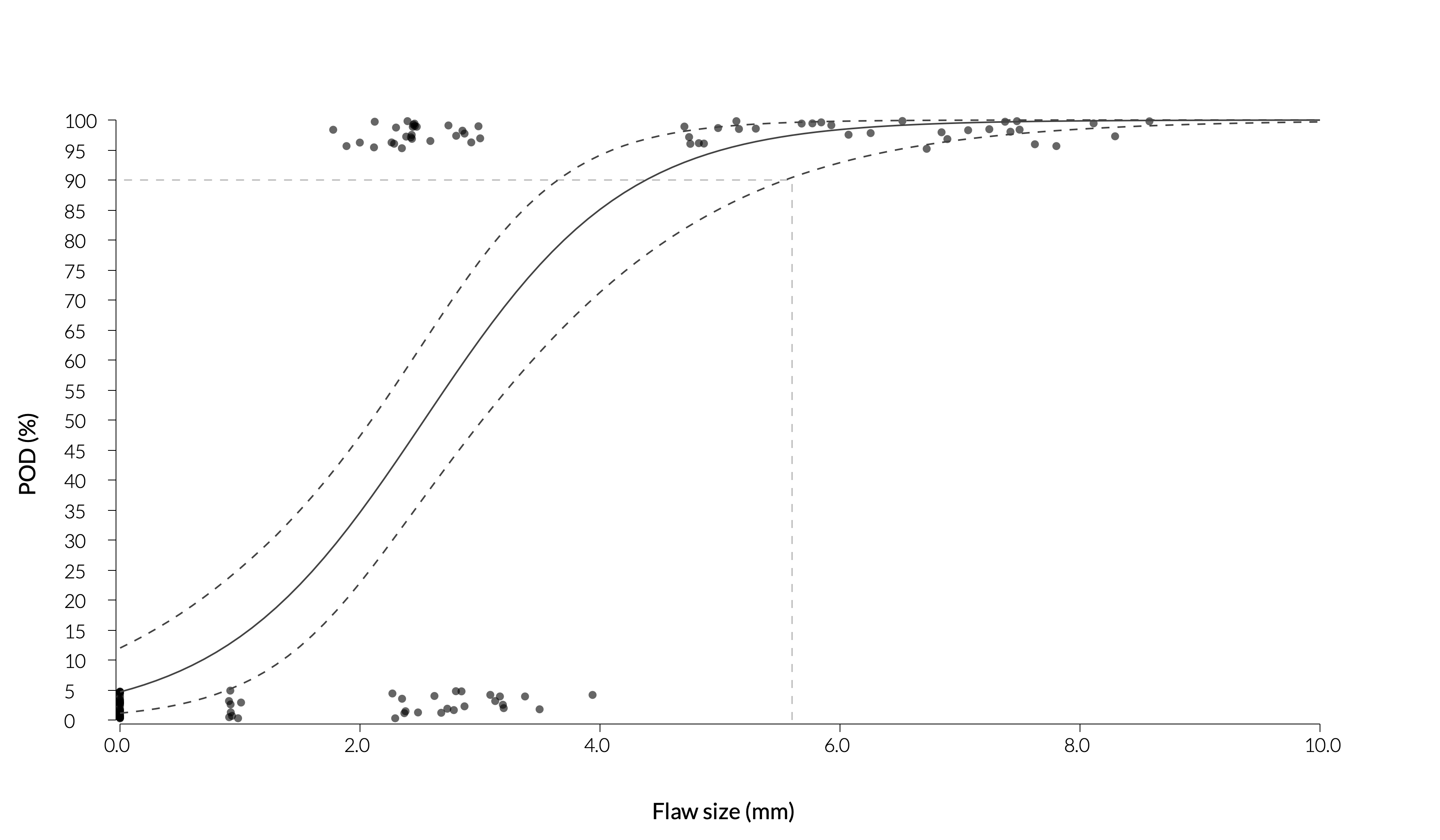}
	\caption{Example POD curve from a human inspector. Note, that additional cracks were added at 0 crack length for comparability on ML-results. The data shows anomalous POD-a dependence due to differences in detectability of various natural cracks. In the future, this can be alleviated by additional cracks to better cover variability in natural cracks.}
	\label{fig:tuhti}
\end{figure}

\begin{figure}
	\includegraphics[width=\linewidth]{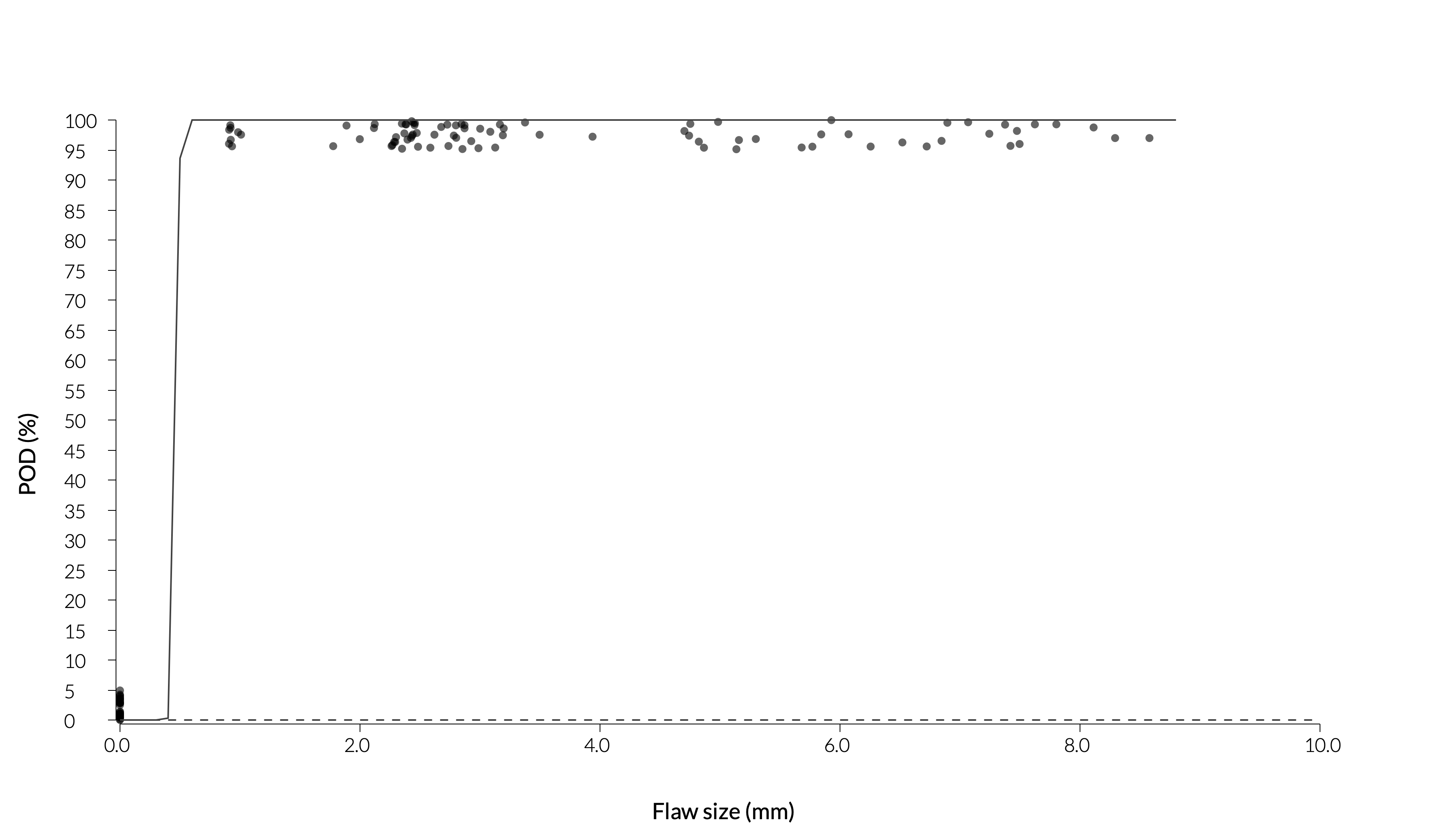}
	\caption{POD curve the machine learning classifier. Note, that additional cracks were added at 0 crack length for convergence.}
	\label{fig:mlpod}
\end{figure}

The performance was evaluated using MIL-HDBK-1823a hit/miss analysis \citep{milhdbk}. The performance comparison is summarized in table \ref{tab:resultcomparison}. POD curve for the human inspectors and the ML network are shown in images \ref{fig:tuhti} and \ref{fig:mlpod}, respectively. As noted in previous research, the cracks contained in the original data presented different challenge in relation to their size. This was primarily caused by the difference in relative amplitude. The same crack was difficult for both the human inspectors and the ML network. In the current data set, the small number of initial flaws as well as their difference caused some irregularities in the hit/miss performance, which the computed confidence bounds to be rather wide. For one inspector, the hits and misses did not show the expected crack size dependence. This may have been caused by excessive false calls for the inspector. For the ML classifier, all the cracks were found. To get convergence for the POD curve, 30 misses of zero-sized cracks were added to all the results. This had the effect of improving slightly the $a_{90/95}$ values of the human inspectors and providing convergence for the ML-classifier even with all the cracks found. In future studies, wider selection of physical cracks are needed to avoid such problems.

\begin{center}
\captionof{table}{Comparison of performance from human inspectors and machine learning classifier. For ML classifier, all the cracks were found and smallest found crack is shown as $a_{90/95}$}. \label{tab:resultcomparison}
\begin{tabular}{ c c c }
\hline
Inspection & $a_{90/50}$ & False calls \\
\hline
Previous data & 1 - 2.5 & 130 \\
\hline
Inspector 1 & 3.0 & 36 \\
Inspector 2 & 2.7 & 917  \\
Inspector 3 & 5.6 & 2    \\
\hline
ML classifier & 0.9 & 0 \\
\hline
\end{tabular}
\end{center}

    \section{Discussion}
The present study showed, that the current very deep machine learning networks are powerful enough to achieve superhuman performance on NDT-tasks previously considered intractable, such as crack detection in ultrasonic signals. This is, to the best of our knowlede, the first time that a direct comparison is published between human inspectors and machine-learning classifiers. Achieving superhuman performance is an important milestone, since it indicates that the machine learning networks can be used also in fields, where high reliability is sought after and regulatory requirements mandate the use of best available means, such as in the nuclear industry.

Data augmentation is a well known technology in the ML literature and is commonly considered to be a key enabling technique when working with limited data sets \citet{deeplearnpyt}. Data augmentation has also previously used for NDT applications of ML \citep{Munir2018}. In present study, extensive data augmentation was utilized using the previously developed virtual flaw technology. This allowed generating training data, that incorporated many aspects of actual inspection, such as the detection of flaw signals from varying backgrounds and variations in probe contact, without extensive data base of real cracks. This can be expected to yield ML-models that generalize well to different real-world inspection cases. In addition, the virtual flaw technology has been used in training human inspectors, and expected to be used in nuclear qualifications in the near future. The use and extensive validation of the virtual flaw technology in the case of human inspectors gives high confidence that the augmented data sets are relevant also for ML applications.

The results from present study indicate, that such domain-specific and separately validated data-augmentation techniques enabling technique for succesfully applying machine learning in various NDE fields, where the data is scarse but performance requirements high.

In previous work, the ML-classification of ultrasonic signal is usually applied at the single A-scan level. In contrast, our approach has been to train the network on full scan of 454 A-scan lines. This approach necessarily limits the applicability of the solution to mechanized or location-encoded inspections, where such coordinated combination of A-scans is available.

The present work has some significant limitations. The raw data contained only three real cracks, that were then modified to give the total data set. This was similar for both the human inspectors and the machine learning solution. The natural flaws exhibit significant variation and a set of three flaws is clearly insufficient to capture this variation. For example, the ASTM POD standard \citep{ASTMahat} requires 30 cracks, which is chiefly to to capture this variation. Thus the network trained here is not expected to work as-is for more general crack detection tasks. Instead, future research will extend the source data using additional thermal fatigue cracks, simulated flaws and other interesting signal types.

    \section{Conclusions}
The following conclusions can be drawn from this study:
    \begin{itemize}
        \item Deep convolutional neural networks are powerful enough to reach superhuman performance in detecting cracks from ultrasonic data
        \item Data augmentation using virtual flaws is seen as key enabling technique to train machine learning networks with limited flawed data
    \end{itemize}

\section{Data availability}
The used python code as well as the training data set is made available for download at \url{https://github.com/iikka-v/ML-NDT}.

\section{Acknowledgements}
The data augmentation using virtual flaws and the initial network \& training was contributed by Trueflaw Ltd. Their contribution is gratefully acknowledged.

\section{References}

\bibliographystyle{apalike}
\bibliography{ndtmlrefs}





\end{document}